\documentclass[aps,prd,onecolumn,floatfix,noshowpacs,tightenlines,noshowkeys,superscriptaddress,amsmath,amssymb,nofootinbib]{revtex4}
\usepackage{amssymb,amsbsy,epsfig,color,graphicx}
\usepackage{color}
\usepackage{[longtable}
\usepackage{array}
\usepackage{dcolumn}   
\usepackage{cellspace}
\usepackage{mathtools}
\usepackage{amstext}
\usepackage{amssymb}
\usepackage{stmaryrd}
\usepackage{stackrel}
\usepackage{graphicx}
\usepackage{esint}
\usepackage[utf8]{inputenc}
\usepackage{blindtext}
\usepackage{float}
\restylefloat{table}
\usepackage{booktabs}
\usepackage{enumitem}
\usepackage{bbold}
\usepackage{slashed}
\usepackage{hyperref}
\usepackage{etoolbox} 
\usepackage{lipsum} 
\usepackage[capitalize]{cleveref}
\usepackage{multirow}
\usepackage[caption=false]{subfig}
\allowdisplaybreaks

\def\be{\begin{equation}}
\def\ee{\end{equation}}
\def\bea{\begin{eqnarray}}
\def\eea{\end{eqnarray}}

\newcommand{\ba}{\begin{eqnarray}}
\newcommand{\ea}{\end{eqnarray}}

\renewcommand\[{\begin{equation}}
\renewcommand\]{\end{equation}}
\makeatletter

\appto{\appendix}{%
\@ifstar{\def\theequation@prefix{A.}}%
{}%
}
\makeatother

\hypersetup{pdfstartview=FitV,colorlinks=true,linkcolor=blue,citecolor=blue,filecolor=black,urlcolor=blue}

\begin{document}

\title{Revealing neutrino nature and $CPT$ violation with decoherence effects}

\author{Luca Buoninfante}
\email{buoninfante.l.aa@m.titech.ac.jp}
\affiliation{Department of Physics, Tokyo Institute of Technology, Tokyo 152-8551, Japan}

\author{Antonio Capolupo}
\email{capolupo@sa.infn.it}
\affiliation{INFN -- Sezione di Napoli, Gruppo collegato di Salerno, I-84084 Fisciano (SA), Italy}
\affiliation{Dipartimento di Fisica "E.R. Caianiello", Universit\`a di Salerno, I-84084 Fisciano (SA), Italy}

\author{Salvatore M. Giampaolo}
\email{sgiampa@irb.hr}
\affiliation{Division of Theoretical Physics, Ruder Bo\v{s}kovi\'{c} Institute, Bijen\v{c}ka cesta 54, 10000 Zagreb, Croatia}

\author{Gaetano Lambiase}
\email{lambiase@sa.infn.it}
\affiliation{INFN -- Sezione di Napoli, Gruppo collegato di Salerno, I-84084 Fisciano (SA), Italy}
\affiliation{Dipartimento di Fisica "E.R. Caianiello", Universit\`a di Salerno, I-84084 Fisciano (SA), Italy}


\begin{abstract}
We study decoherence effects on mixing among three generations of neutrinos. We show that in presence of a non--diagonal dissipation matrix, both Dirac and Majorana neutrinos can violate the $CPT$ symmetry and the oscillation formulae depend on the parametrization of the mixing matrix. We reveal the $CP$ violation in the transitions preserving the flavor, for a certain form of the dissipator. In particular, the $CP$ violation affects all the transitions in the case of Majorana neutrinos, unlike   Dirac neutrinos
  which still preserve the $CP$ symmetry in one of the transitions flavor preserving.
This theoretical result shows that decoherence effects, if exist for neutrinos, could allow to determine the neutrino nature and to test fundamental symmetries of physics.
Next long baseline experiments could allow such an analysis.
  We relate our study with experiments by using the characteristic parameters and the constraints on the elements of the dissipation matrix of current experiments.
\end{abstract}

\maketitle


\section{Introduction}

Nowadays the concept of neutrino mixing/oscillation represent one of the main missing ingredient in the Standard Model of particles, indeed its experimental verifications \cite{Nakamura1,Nakamura2,Nakamura3,Nakamura4,Nakamura5,Nakamura6} stimulated lots of new investigations aimed to extend the standard theory by including a non--zero mass for neutrinos. One of the most important open issues, at both theoretical and experimental levels, is to determine the values of neutrino  masses and to understand their real nature, i.e. whether they are Dirac or Majorana particles.

The most known and studied physical effect which could shed some light on neutrino nature is the neutrinoless double beta decay for which several experiments have been proposed \cite{Giuliani}, but so far no results have been obtained. Recently, to discriminate between Dirac and Majorana neutrinos also other scenarios have been proposed in which the fundamental physical quantity is not the decay rate of a process but, for instance, the Leggett--Garg $K_{3}$ quantity \cite{Richter} and  the geometric phase for neutrinos  \cite{Capolupo2018}. Moreover, it is also well known that the neutrino oscillation formulae in the presence of decoherence can depend on the Majorana phase \cite{Benatti}. This feature was used by the authors in Ref.\cite{Capolupo:2018hrp} in the case of two flavors neutrinos to explicitly show how an off--diagonal dissipator can distinguish between the two kind of neutrinos and that one of the physical implications is the violation of $CPT$ symmetry.

According to the $CPT$ theorem, the Hamiltonian of a Lorentz invariant local quantum field theory is invariant under a simultaneous transformation of charge conjugation $C$, parity inversion $P$ and time reversal $T$, so that $CPT$ turns out to be an exact fundamental symmetry \cite {Kost}. However, such a theorem is based on the crucial assumption that any kind of decoherence and dissipation effects are negligible.

The phenomena of dissipation and decoherence could be consequences of the interaction between neutrinos and the surrounding environment, or space--time fluctuations induced by quantum gravity effects. Many efforts have been already made in
order the study dissipation and its origin  in neutrino oscillations \cite{Benatti,BenattiB,Benatti1,Simonov:2019mvt}.

Here, we  extend the study performed in \cite{Capolupo:2018hrp} to the case of three flavors neutrinos and we reveal new features due to the presence of Dirac and Majorana phases in the mixing matrix. We consider diagonal and off--diagonal dissipators and  we analyze the time evolution of the density matrix for neutrinos.
We show that for an off--diagonal dissipator, in the three flavor mixing case, $CPT$ symmetry can be broken both for Dirac and Majorana neutrinos because of the presence of different  phases in the mixing matrix. This result is different with respect to that obtained in the case of two flavor mixing  for which $CPT$ symmetry is violated only by Majorana neutrinos \cite{Capolupo:2018hrp}.  Another characteristic behavior of the mixing among three families here revealed is that, for a simple off--diagonal dissipator,  Majorana neutrinos can violate $CP$ symmetry in all the flavor preserving neutrino transitions because of the presence of three phases (the Dirac phase and the two Majorana phases) in the mixing matrix. On the contrary, Dirac neutrinos can break $CP$ symmetry only in two of the three flavor preserving transitions.
A difference between Dirac and Majorana neutrinos can be revealed also in the case of diagonal dissipator with  $\gamma_1 \neq \gamma_2$, or $ \gamma_4 \neq \gamma_5$, or $ \gamma_6 \neq \gamma_7$.
Moreover,  we show that the oscillation formulae for Majorana neutrinos depend on the  parametrization of the Majorana mixing matrix. Therefore, if the decoherence affects neutrino evolution, the oscillation formulae could reveal the neutrino nature and, if the neutrinos are Majorana fermions, one could determine the right parametrization of the Majorana mixing matrix. Our theoretical studies suggest that the neutrino nature and the violation of fundamental symmetries could be analyzed with future long baseline experiments \cite{Aartsen:2017nmd,DUNE}. We consider the neutrino oscillations in vacuum since in this case the  violation of $CP$ and $CPT$ symmetries due to the decoherence  are not affected by other phenomena.
In  fact, for neutrinos travelling, for example, through Earth,
the MSW effect  already introduces an additional degree of $CPT$ violation \cite{PDG}. Therefore, one has to be careful to identify the right contribution responsible for violations purely induced
by decoherence. Since we are mainly interested in highlighting the difference between Dirac and Majorana neutrinos,
for simplicity we compare them in the vacuum. In a forthcoming paper we will extend our treatment in the
presence of matter.

The paper is organized as follows. In Section \ref{dirac-majorana} we briefly review the concepts of Dirac and Majorana neutrinos and introduce the mathematical tools of the density matrix needed to compute all the oscillation formulae for three flavors neutrinos in presence of decoherence. In Section \ref{diagon-diss}, we consider a diagonal dissipator and show  that in this case the oscillation formulae are independent of the neutrino nature. In Section \ref{non-diag dissip}, we show the effects of an off--diagonal dissipator on the oscillation formulae and
 on the violation of $CP$ and $CPT$ symmetries. Moreover, we show the dependence of these quantities on the representation of the Majorana mixing matrix.
In Section \ref{numeric-analysis}, we present a numerical analysis by using the available data of the characteristic parameters involved in long baseline experiments. In Section \ref{conclus} we summarize the contents of this paper by emphasizing the relevance of the main results, and draw our conclusions.

\section{Neutrino mixing and decoherence}\label{dirac-majorana}

The main distinction between Dirac and Majorana neutrinos relies on the fact that Dirac Lagrangian is invariant under the global transformation of $U(1)$ so that all the associated charges (like electric, leptonic, etc.) turn out to be conserved, while Majorana Lagrangian breaks the $U(1)$  symmetry. A process in which the lepton number is violated and therefore would be allowed only for Majorana neutrinos and not for Dirac is the neutrinoless double beta decay.

The breaking of the $U(1)$ global symmetry has also consequences on the form of the mixing matrix \cite{Maki:1962mu} which contains a different number of physical phases for the two kind of neutrinos. Indeed, in the general case of the mixing with $n$ Dirac fields,  there exist $N_{D} = \frac{(n -1)(n-2)}{2}$ physical phases, while for $n$  Majorana fields, one has additional $N_{M} = \frac{n (n - 1) }{2}$ phases. The $n-1$ extra phases are called Majorana phases and their detection would allow to identify the nature of neutrinos.

Let us recall that the Lagrangian density for Dirac neutrinos in flavor basis is given by
\begin{equation}
\mathcal{L}(x)=\bar{\Psi}_f(x)\left(i\slashed{\partial}-M\right)\Psi_f(x),\label{dirac-lagran}
\end{equation}
where $\Psi^{T}_f=(\nu_e,\nu_{\mu},\nu_{\tau})$ and $M^{\dagger}=M$ is the mixed mass term. The mixing relations are
\cite{Maki:1962mu,ref-ckm}:
\begin{equation}
\Psi_f(x)=\mathcal{U}_D\Psi_{m}(x)=\left(\begin{matrix}
c_{12}c_{13}& s_{12}c_{13} & s_{13}e^{-i\delta}\\
-s_{12}c_{23}-c_{12}s_{23}s_{13}e^{i\delta}& c_{12}c_{23}-s_{12}s_{23}s_{13}e^{i\delta} &s_{23}c_{13}\\
s_{12}s_{23}-c_{12}c_{23}s_{13}e^{i\delta}& -c_{12}s_{23}-s_{12}c_{23}s_{13}e^{i\delta} & c_{23}c_{13}
\end{matrix}\right)\Psi_m(x),\label{CKM}
\end{equation}
where $\mathcal{U}_D$ is the Dirac mixing matrix, $\delta$ is the  Dirac phase, $c_{ij}={\rm cos}(\theta_{ij})$ and $s_{ij}={\rm sin}(\theta_{ij})\,,$ with $\theta_{ij}$ being the  mixing angles between the fields with definite masses $\nu_i,\,\nu_j$ with $i,j=e,\mu,\tau\,,$ and $\Psi^{T}_m=(\nu_1,\nu_2,\nu_3)\,.$
 Eq.\eqref{dirac-lagran} is diagonalized  by using Eq.\eqref{CKM}, so that we obtain the Lagrangian for free Dirac fermions with masses $m_1,$ $m_2$ and $m_3 :$
\begin{equation}
\mathcal{L}(x)=\bar{\Psi}_m(x)\left(i\slashed{\partial}-M_d\right)\Psi_m(x) \label{dirac-lagran-diag},
\end{equation}
where $M_d={\rm diag}(m_1,m_2,m_3)\,.$

For Majorana neutrinos, different parametrizations of the mixing matrix $\mathcal{U}_M,$  exist.
 In fact, when decoherence is negligible, and even in the case of a diagonal dissipator, all the transition probabilities turn out to be invariant under the rephasing $\mathcal{U}_{\alpha k} \rightarrow e^{i \phi_{k}} \mathcal{U}_{\alpha k} $  $(\alpha = e, \mu; k =1,2)$. This means that the Majorana phases $\phi_{i}$  do not affect the oscillation formulae which are the same as for Dirac neutrinos \cite{Pontecorvo}.
 For instance, one can write
\begin{equation}
\mathcal{U}_M=\mathcal{U}_D\cdot {\rm diag}\left(1,e^{i\phi_1},e^{i\phi_2}\right),\label{non-interest-majo}
\end{equation}
where $\phi_1$ and $\phi_2$ are the two Majorana phases. Another possible parametrization is the following:
\begin{equation}
\begin{array}{rl}
\mathcal{U}_M=&\displaystyle \left(1,e^{-i\phi_1},e^{-i\phi_2}\right)\cdot \mathcal{U}_D\cdot {\rm diag}\left(1,e^{i\phi_1},e^{i\phi_2}\right)\\[3mm]
=& \displaystyle \left(\begin{matrix}
c_{12}c_{13}& s_{12}c_{13}e^{i\phi_1} & s_{13}e^{i(\phi_2-\delta)}\\
-s_{12}c_{23}e^{-\phi_1}-c_{12}s_{23}s_{13}e^{i(\delta-\phi_1)}& c_{12}c_{23}-s_{12}s_{23}s_{13}e^{i\delta} &s_{23}c_{13}e^{i(\phi_2-\phi_1)}\\
s_{12}s_{23}e^{-i\phi_2}-c_{12}c_{23}s_{13}e^{i(\delta-\phi_2)}& -c_{12}s_{23}e^{i(\phi_1-\phi_2)}-s_{12}c_{23}s_{13}e^{i(\delta+\phi_1-\phi_2)} & c_{23}c_{13}
\end{matrix}\right)\,,
\end{array}
\label{majorana-matrix}
\end{equation}
and other choices leading to the same oscillation formulae are presented in Ref.\cite{Giunti}.

This fact is no longer true when there are off--diagonal elements in dissipation matrix
and also in the case of diagonal dissipator with  $\gamma_1 \neq \gamma_2$, or $ \gamma_4 \neq \gamma_5$, or $ \gamma_6 \neq \gamma_7$.
Indeed, one can obtain oscillation formulae for Majorana neutrinos   depending on the phases $\phi_{i}$, and  on the   parametrization of the mixing matrix, as shown in Ref.\cite{Capolupo:2018hrp} for two flavor mixing and non--diagonal dissipator.
In the following we will consider the case of three flavor neutrino mixing and reveal new aspects of neutrino oscillations which are absent in the case of mixing between two neutrinos.
In the rest of the paper, we mainly focus on the matrix given in \eqref{majorana-matrix}, which will be very useful to highlight the main features in presence of decoherence.

By treating the neutrino as an open quantum system, we analyze the physical implications of decoherence in flavor mixing.
In particular, we study the time evolution of the density matrix corresponding to the neutrino state in the flavor basis and compute several transition probabilities for both diagonal and non--diagonal dissipation matrix.

The state evolution of neutrinos seen as an open system, can be described by the Lindblad-Kossakowski master equation \cite{Lind}:
\begin{equation}
\frac{\partial \rho(t)}{\partial t}=-i\left[H,\rho(t)\right]+D[\rho(t)],\label{lindbland}
\end{equation}
where $H=H^{\dagger}$ is the total Hamiltonian of the system and $D[\rho(t)]$ is the dissipator defined as
\begin{equation}
D[\rho(t)]=\frac{1}{2}\sum_{i,j=0}^{N^2-1}a_{ij}\left( \left[F_i\rho(t),F_j^{\dagger}\right]+\left[F_i,\rho(t)F_j^{\dagger}\right] \right)\,,\label{dissip}
\end{equation}
with $a_{ij}$  Kossakowski coefficients whose form is related to the characteristics of the environment \cite{Benatti}. The operators $F_i\,,$ with $i=1,\dots, N^2-1\,,$  satisfy the relations ${\rm Tr}(F_i)=0$ and ${\rm Tr}\left(F_i^{\dagger}F_j\right)=\delta_{ij},$ and in the case of three flavor neutrinos they are the Gell-Mann matrices $\lambda_i$ which satisfy the following properties:
\begin{equation}
\lambda^{\dagger}_i=\lambda_i,\quad [\lambda_i,\lambda_j]=2if_{ijk}\lambda_k,\quad f_{ijk}=-\frac{i}{4}{\rm Tr}\left(\lambda_i[\lambda_j,\lambda_k]\right).
\end{equation}
Here the non--vanishing $f_{ijk}$ are given by $f^{123}=1,$ $f^{147}=f^{165}=f^{246}=f^{257}=f^{345}=f^{376}=\frac{1}{2},$ $f^{458}=f^{678}=\frac{\sqrt{3}}{2}\,.$

Let us now expand Eqs.(\ref{lindbland}) and (\ref{dissip}) in the basis of $SU(3):$
\begin{equation}
\dot{\rho}_{\mu}(t)=f_{ij\mu}H_i\rho_j(t)+D_{\mu\nu} \rho_{\nu}(t)\,,\label{exp-su(3)}
\end{equation}
where $\rho_{\mu}={\rm Tr}\left(\rho \lambda_{\mu}\right),$ with $\mu=0,\dots,8\,.$  Given the mass differences $\Delta m_{21}^2=m_2^2-m_1^2$ and $\Delta m_{31}^2=m_3^2-m_1^2,$ the Hamiltonian reads
\begin{equation}
H=\frac{1}{2E}\left(\begin{matrix}
0&0 &0\\
0& \Delta m_{21}^2 &0\\
0&0 &\Delta m_{31}^2
\end{matrix}\right)\equiv \displaystyle  \left(\begin{matrix}
0&0 &0\\
0& \Delta_{21} &0\\
0&0 &\Delta_{31}
\end{matrix}\right),\label{hamilt-matrix}
\end{equation}
where  $\Delta_{21}=\frac{1}{2E}\Delta m_{21}^2$ and $\Delta_{31}=\frac{1}{2E}\Delta m_{31}^2\,.$ One can show that the only non--vanishing components $H_\mu$ are
\begin{equation}
H_0=\Delta_{21}+\Delta_{31},\quad H_3=-\Delta_{21},\quad H_{8}=\frac{1}{\sqrt{3}}\left(\Delta_{21}-2\Delta_{31}\right)\,.\label{non-vanish-compon}
\end{equation}
The dissipator in Eq.(\ref{exp-su(3)}) is given by
\begin{equation}
D_{\mu\nu}=-\left(\begin{matrix}
0 & 0 & 0 & 0 & 0 & 0 & 0 & 0 & 0 \\
0 & \gamma_1 & \alpha_1 &\beta_1 &\delta_1 & \chi_1  &\xi_1&\zeta_1& \eta_1 \\
0 & \alpha_1 & \gamma_2 &\alpha_2 & \beta_2&\delta_2 &\chi_2 &\xi_2 & \zeta_2\\
0 & \beta_1 & \alpha_2& \gamma_3 & \alpha_3&\beta_3 &\delta_3 &\chi_3 & \xi_3\\
0 & \delta_1 &\beta_2 & \alpha_3&\gamma_4 &\alpha_4 &\beta_4 &\delta_4 & \chi_4 \\
0 & \chi_1 &\delta_2 & \beta_3& \alpha_4&\gamma_5 &\alpha_5 & \beta_5&\delta_5 \\
0 & \xi_1 &\chi_2 &\delta_3 &\beta_4 &\alpha_5 & \gamma_6&\alpha_6 &\beta_6 \\
0 & \zeta_1 &\xi_2 &\chi_3 &\delta_4 &\beta_5 &\alpha_6 &\gamma_7 & \alpha_7\\
0 & \eta_1 &\zeta_2 &\xi_3 &\chi_4 &\delta_5 & \beta_6&\alpha_7 &\gamma_8
\end{matrix}\right)\,.\label{dissip-matrix}
\end{equation}
where we considered the probability conservation which implies $D_{\mu 0}=D_{0\nu}=0.$
All the elements in the matrix \eqref{dissip-matrix} are real and the ones on the diagonal are positive  in order to satisfy the relation ${\rm Tr}\left(\rho(t)\right)=1\,.$ Hence, from Eq.\eqref{exp-su(3)} it is now clear that we have nine equations among which the $\mu=0$ component is trivial. Indeed, since $f_{ij0}=0$ and $D_{0\nu}=0$ we obtain $\dot{\rho}_0(t)=0\Rightarrow \rho_0(t)=1.$

The density matrix written in terms of the components $\rho_{\mu}$ in the basis $\lambda_{\mu}$ reads
\begin{equation}
\begin{array}{rl}
\rho(t)=&\displaystyle \frac{1}{3}\rho_0(t)\lambda_0+\frac{1}{2}\sum\limits_{i=1}^{8}\rho_i(t)\lambda_i\\[4mm]
=&\displaystyle\frac{1}{2}\left(\begin{matrix}\displaystyle
\frac{2}{3}\rho_0+\rho_3+\frac{\rho_8}{\sqrt{3}}& \rho_1-i\rho_2 & \rho_4-i\rho_5\\
\rho_1+i\rho_2& \displaystyle\frac{2}{3}\rho_0-\rho_3+\frac{\rho_8}{\sqrt{3}}&\rho_6-i\rho_7\\
\rho_4+i\rho_5& \rho_6+i\rho_7&\displaystyle\frac{2}{3}\rho_0-\frac{2}{\sqrt{3}}\rho_8
\end{matrix}\right)\,.\label{density matrix}
\end{array}
\end{equation}
With this expression of the density matrix, the neutrino oscillation formulae reads
\bea
P_{\nu_a \rightarrow \nu_b} = \frac{1}{3} + \frac{1}{2}\sum_{i = 1}^{8} \rho_{a,i}(t) \rho_{b,i}(0).
\eea
Notice that, the $CP$ symmetry violation is defined as $\Delta CP_{ab}\equiv P_{\nu_a \rightarrow \nu_b}-  P_{\bar{\nu}_a \rightarrow \bar{\nu}_b} \neq 0$ and
the $T$ violation is given by $\Delta T_{ab}\equiv P_{\nu_a \rightarrow \nu_b} -  P_{ \nu_b \rightarrow  \nu_a}\neq 0.$ The $CPT$ symmetry is violated when $\Delta CP \neq \Delta T$.

\section{Diagonal dissipator}\label{diagon-diss}

We now study decoherence effects considering the mixing matrix  \eqref{majorana-matrix}. We analyze both cases of zero and non--zero Majorana phases. We start by solving the set of equations \eqref{exp-su(3)} in the simpler case of a diagonal dissipator:
\begin{equation}
D_{\mu\nu}=-{\rm diag}\left(0,\gamma_1,\gamma_2,\gamma_3,\gamma_4,\gamma_5,\gamma_6,\gamma_7,\gamma_8\right)\,.\label{diag-dissip}
\end{equation}
Then, the system of differential equations \eqref{exp-su(3)} becomes
\begin{eqnarray}
\dot{\rho_0}(t)=\nonumber&&0\,,\\[3mm]\nonumber
\dot{\rho}_1(t)=&&\displaystyle \Delta_{21}\rho_2(t)-\gamma_1 \rho_1(t)\,, \\[3mm]\nonumber
\dot{\rho}_2(t)=&&\displaystyle  -\Delta_{21}\rho_1(t)-\gamma_2 \rho_2(t)\,, \\[3mm]\nonumber
\dot{\rho}_3(t)=&& \displaystyle - \gamma_3 \rho_3 (t)\,,\\[3mm]\nonumber
\dot{\rho}_4(t)=&&\displaystyle  \Delta_{31}\rho_5(t)-\gamma_4 \rho_4(t)\,, \\[3mm]\nonumber
\dot{\rho}_5(t)=&&\displaystyle  -\Delta_{31}\rho_4(t)-\gamma_5 \rho_5(t)\,, \\[3mm]\nonumber
\dot{\rho}_6(t)=&&\displaystyle  \Delta_{32}\rho_7(t)-\gamma_6 \rho_6(t)\,, \\[3mm]\nonumber
\dot{\rho}_7(t)=&&\displaystyle-\Delta_{32}\rho_6(t)-\gamma_7 \rho_7(t)\,,  \\[3mm]
\dot{\rho}_8(t)=&&\displaystyle -\gamma_8 \rho_8(t)\,,
\label{syst-eq-diag}
\end{eqnarray}
where $\Delta_{32}=\Delta_{31}-\Delta_{21}=\frac{\Delta m_{32}^2}{2E}.$

We consider now the diagonal dissipator Eq.(\ref{diag-dissip}) with  the conditions: $\gamma_1 = \gamma_2 = \gamma_{12}$, $\gamma_4= \gamma_5 = \gamma_{45}$, $\gamma_6= \gamma_7 = \gamma_{67} $. This choice is consistent with that of ref.\cite{Balieiro}.
 The system of equations can be solved as follows:
\begin{eqnarray}
\rho_{0}(t)=\nonumber&&\displaystyle 1,\\[3mm]\nonumber
\rho_{1}(t)=&&\displaystyle e^{-\gamma_{12} t}\left[\rho_1(0){\rm cos}(\Delta_{21}t)+\rho_{2}(0){\rm sin}(\Delta_{21}t)\right]\,, \\[3mm]\nonumber
\rho_{2}(t)=&&\displaystyle  e^{-\gamma_{12} t}\left[\rho_2(0){\rm cos}(\Delta_{21}t)-\rho_{1}(0){\rm sin}(\Delta_{21}t)\right]\,, \\[3mm]\nonumber
\rho_{3}(t)=&& \displaystyle e^{-\gamma_{3} t} \rho_3(0)\,,\\[3mm]\nonumber
\rho_{4}(t)=&&\displaystyle  e^{-\gamma_{45} t}\left[\rho_4(0){\rm cos}(\Delta_{31}t)+\rho_{5}(0){\rm sin}(\Delta_{31}t)\right]\,, \\[3mm]\nonumber
\rho_{5}(t)=&&\displaystyle  e^{-\gamma_{45} t}\left[\rho_5(0){\rm cos}(\Delta_{31}t)-\rho_{4}(0){\rm sin}(\Delta_{31}t)\right]\,, \\[3mm]\nonumber
\rho_{6}(t)=&&\displaystyle  e^{-\gamma_{67} t}\left[\rho_6(0){\rm cos}(\Delta_{32}t)+\rho_{7}(0){\rm sin}(\Delta_{32}t)\right]\,, \\[3mm]\nonumber
\rho_{7}(t)=&&e^{-\gamma_{67} t}\left[\rho_7(0){\rm cos}(\Delta_{32}t)-\rho_{6}(0){\rm sin}(\Delta_{32}t)\right]\,, \\[3mm]
\rho_{8}(t)=&&\displaystyle e^{-\gamma_8 t} \rho_8(0)\,.
\label{sol-diag}
\end{eqnarray}
The initial conditions $\rho_i(0)$ can be found by employing the following relations:
$
\rho_{a}(0)=\left|\nu_a \right\rangle  \left\langle\nu_a \right| \,,\quad a=e,\mu,\tau\,.
$
For electronic neutrino we have
\begin{eqnarray}
\rho_{e,0}(0)=\nonumber&&\displaystyle 1,\\[3mm]\nonumber
\rho_{e,1}(0)=&&\displaystyle {\rm sin}(2\theta_{12}){\rm cos}^2\theta_{13}\,{\rm cos}\,\phi_1\,, \\[3mm]\nonumber
\rho_{e,2}(0)=&&\displaystyle  {\rm sin}(2\theta_{12}){\rm cos}^2\theta_{13}\,{\rm sin}\,\phi_1\,, \\[3mm]\nonumber
\rho_{e,3}(0)=&& \displaystyle {\rm cos}^2\theta_{13}\left(2{\rm cos}^2\theta_{12}-1\right)\,,\\[3mm]\nonumber
\rho_{e,4}(0)=&&\displaystyle  {\rm sin}(2\theta_{13}){\rm cos}\,\theta_{12}{\rm cos}(\phi_2-\delta)\,, \\[3mm]\nonumber
\rho_{e,5}(0)=&&\displaystyle  {\rm sin}(2\theta_{13}){\rm cos}\,\theta_{12}{\rm sin}(\phi_2-\delta)\,, \\[3mm]\nonumber
\rho_{e,6}(0)=&&\displaystyle  {\rm sin}(2\theta_{13}){\rm sin}\,\theta_{12}{\rm cos}(\phi_2-\phi_1-\delta)\,, \\[3mm]\nonumber
\rho_{e,7}(0)=&&\displaystyle {\rm sin}(2\theta_{13}){\rm sin}\,\theta_{12}{\rm sin}(\phi_2-\phi_1-\delta)\,, \\[3mm]
\rho_{e,8}(0)=&&\displaystyle \sqrt{3}\left(\frac{1}{3}-{\rm sin}^2\theta_{13}\right)\,.
\label{sol-diag-electronic}
\end{eqnarray}
For muon neutrino we obtain
\begin{eqnarray}
\rho_{\mu,0}(0)=\nonumber&&\displaystyle 1,\\[3mm]\nonumber
\rho_{\mu,1}(0)=&&\displaystyle -{\rm sin}(2\theta_{12}){\rm cos}^2\theta_{23}{\rm cos}\,\phi_1-{\rm sin}(2\theta_{23}){\rm sin}\,\theta_{13}{\rm cos}^2\theta_{12}{\rm cos}(\delta-\phi_1) \\[3mm]\nonumber
&&\displaystyle +{\rm sin}(2\theta_{23}){\rm sin}^2\theta_{12}{\rm sin}\,\theta_{13}{\rm cos}(\delta+\phi_1)+{\rm sin}(2\theta_{12}){\rm sin}^2\theta_{23}{\rm sin}^2\theta_{13}{\rm cos}\rm \phi_1\,,\\[3mm]\nonumber
\rho_{\mu,2}(0)=&&\displaystyle -{\rm sin}(2\theta_{12}){\rm cos}^2\theta_{23}{\rm sin}\,\phi_1+{\rm sin}(2\theta_{23}){\rm sin}\,\theta_{13}{\rm cos}^2\theta_{12}{\rm sin}(\delta-\phi_1) \\[3mm]\nonumber
&&\displaystyle +{\rm sin}(2\theta_{23}){\rm sin}^2\theta_{12}{\rm sin}\,\theta_{13}{\rm sin}(\delta+\phi_1)+{\rm sin}(2\theta_{12}){\rm sin}^2\theta_{23}{\rm sin}^2\theta_{13}{\rm sin}\rm \phi_1\,,\\[3mm]\nonumber
\rho_{\mu,3}(0)=&& \displaystyle -1+{\rm sin}^2\theta_{23}{\rm cos}^2\theta_{13}+2{\rm sin}^2\theta_{12}{\rm cos}^2\theta_{23}+2{\rm cos}^2\theta_{12}{\rm sin}^2\theta_{23}{\rm sin}^2\theta_{13}\\[3mm]\nonumber
&&\displaystyle +{\rm sin}(2\theta_{12}){\rm sin}(2\theta_{23}){\rm sin}\,\theta_{13}{\rm cos}\,\delta\,,\\[3mm]\nonumber
\rho_{\mu,4}(0)=&&\displaystyle -{\rm sin}(2\theta_{23}){\rm sin}\,\theta_{12}{\rm cos}\,\theta_{13}{\rm cos}\,\phi_2-{\rm sin}(2\theta_{13}){\rm cos}\,\theta_{12}{\rm sin}^2\theta_{23}{\rm cos}(\phi_2-\delta) \,, \\[3mm]\nonumber
\rho_{\mu,5}(0)=&&\displaystyle   -{\rm sin}(2\theta_{23}){\rm sin}\,\theta_{12}{\rm cos}\,\theta_{13}{\rm sin}\,\phi_2-{\rm sin}(2\theta_{13}){\rm cos}\,\theta_{12}{\rm sin}^2\theta_{23}{\rm sin}(\phi_2-\delta) \,, \\[3mm]\nonumber
\rho_{\mu,6}(0)=&&\displaystyle   {\rm sin}(2\theta_{23}){\rm cos}\,\theta_{12}{\rm cos}\,\theta_{13}{\rm cos}(\phi_2-\phi_1)-{\rm sin}(2\theta_{13}){\rm sin}\,\theta_{12}{\rm sin}^2\theta_{23}{\rm cos}(\phi_2-\phi_1-\delta) \,, \\[3mm]\nonumber
\rho_{\mu,7}(0)=&&\displaystyle {\rm sin}(2\theta_{23}){\rm cos}\,\theta_{12}{\rm cos}\,\theta_{13}{\rm sin}(\phi_2-\phi_1)-{\rm sin}(2\theta_{13}){\rm sin}\,\theta_{12}{\rm sin}^2\theta_{23}{\rm sin}(\phi_2-\phi_1-\delta)\,, \\[3mm]
\rho_{\mu,8}(0)=&&\displaystyle \sqrt{3}\left(\frac{1}{3}-{\rm sin}^2\theta_{23}{\rm cos}^2\theta_{13}\right)\,;
\label{sol-diag-muonic}
\end{eqnarray}
and finally for tau neutrino
\begin{eqnarray}
\rho_{\tau,0}(0)=\nonumber&&\displaystyle 1,\\[3mm]\nonumber
\rho_{\tau,1}(0)=&&\displaystyle -{\rm sin}(2\theta_{12}){\rm sin}^2\theta_{23}{\rm cos}\,\phi_1+{\rm sin}(2\theta_{23}){\rm sin}\,\theta_{13}{\rm cos}^2\theta_{12}{\rm cos}(\phi_1-\delta) \\[3mm]\nonumber
&&\displaystyle -{\rm sin}(2\theta_{23}){\rm sin}^2\theta_{12}{\rm sin}\,\theta_{13}{\rm cos}(\delta+\phi_1)+{\rm sin}(2\theta_{12}){\rm cos}^2\theta_{23}{\rm sin}^2\theta_{13}{\rm cos}\rm \phi_1\,,\\[3mm]\nonumber
\rho_{\tau,2}(0)=&&\displaystyle -{\rm sin}(2\theta_{12}){\rm sin}^2\theta_{23}{\rm sin}\,\phi_1+{\rm sin}(2\theta_{23}){\rm sin}\,\theta_{13}{\rm cos}^2\theta_{12}{\rm sin}(\phi_1-\delta) \\[3mm]\nonumber
&&\displaystyle -{\rm sin}(2\theta_{23}){\rm sin}^2\theta_{12}{\rm sin}\,\theta_{13}{\rm cos}(\delta+\phi_1)+{\rm sin}(2\theta_{12}){\rm cos}^2\theta_{23}{\rm sin}^2\theta_{13}{\rm cos}\rm \phi_1\,,\\[3mm]\nonumber
\rho_{\tau,3}(0)=&& \displaystyle -1+{\rm cos}^2\theta_{23}{\rm cos}^2\theta_{13}+2{\rm sin}^2\theta_{12}{\rm sin}^2\theta_{23}+2{\rm cos}^2\theta_{12}{\rm cos}^2\theta_{23}{\rm sin}^2\theta_{13}\\[3mm]\nonumber
&&\displaystyle -{\rm sin}(2\theta_{12}){\rm sin}(2\theta_{23}){\rm sin}\,\theta_{13}{\rm cos}\,\delta\,,\\[3mm]\nonumber
\rho_{\tau,4}(0)=&&\displaystyle {\rm sin}(2\theta_{23}){\rm sin}\,\theta_{12}{\rm cos}\,\theta_{13}{\rm cos}\,\phi_2-{\rm sin}(2\theta_{13}){\rm cos}\,\theta_{12}{\rm cos}^2\theta_{23}{\rm cos}(\phi_2-\delta) \,, \\[3mm]\nonumber
\rho_{\tau,5}(0)=&&\displaystyle {\rm sin}(2\theta_{23}){\rm sin}\,\theta_{12}{\rm cos}\,\theta_{13}{\rm sin}\,\phi_2-{\rm sin}(2\theta_{13}){\rm cos}\,\theta_{12}{\rm cos}^2\theta_{23}{\rm sin}(\phi_2-\delta) \,, \\[3mm]\nonumber
\rho_{\tau,6}(0)=&&\displaystyle   -{\rm sin}(2\theta_{23}){\rm cos}\,\theta_{12}{\rm cos}\,\theta_{13}{\rm cos}(\phi_1-\phi_2)-{\rm sin}(2\theta_{13}){\rm sin}\,\theta_{12}{\rm cos}^2\theta_{23}{\rm cos}(\delta+\phi_1-\phi_2) \,, \\[3mm]\nonumber
\rho_{\tau,7}(0)=&&\displaystyle {\rm sin}(2\theta_{23}){\rm cos}\,\theta_{12}{\rm cos}\,\theta_{13}{\rm sin}(\phi_1-\phi_2)+{\rm sin}(2\theta_{13}){\rm sin}\,\theta_{12}{\rm cos}^2\theta_{23}{\rm sin}(\delta+\phi_1-\phi_2) \,, \\[3mm]
\rho_{\tau,8}(0)=&&\displaystyle \sqrt{3}\left(\frac{1}{3}-{\rm cos}^2\theta_{23}{\rm cos}^2\theta_{13}\right)\,.
\label{sol-diag-tau}
\end{eqnarray}
The neutrino oscillation  probabilities, as said above, are obtained through the relation $P_{\nu_a\rightarrow\nu_b}={\rm Tr}\left[\rho_b(t)\cdot \rho_a(0)\right]\,.$ By  computing the transition probability in the case of a diagonal dissipator as in Eq.\eqref{diag-dissip}, for flavor preserving transitions we obtain
\begin{equation}
\Delta CP_{aa} = P_{\nu_a\rightarrow \nu_a}-P_{\bar{\nu}_{a}\rightarrow \bar{\nu}_a} = 0\,,\quad a=e,\mu,\tau\,\\[3mm].
\end{equation}
In similar way, $\Delta T_{aa} = 0$. These result are the same of those obtained in the absence of decoherence. Moreover, like in the standard case, $CP$ and $T$ symmetries are violated because of the presence of the Dirac phase $\delta,$ while the presence of diagonal elements in the dissipation matrix only introduces a damping factor which is physically expected. For instance, the three channels responsible for $CP$ violations read
\begin{equation}
\begin{array}{rl}
\Delta CP_{e\mu}=&P_{\nu_{e}\rightarrow \nu_{\mu}}-P_{\bar{\nu}_{e}\rightarrow \bar{\nu}_{\mu}} \\[3mm]
=&   \displaystyle {\rm sin}\,\delta\, {\rm cos}^2\theta_{13}{\rm sin}(\theta_{12}){\rm sin}(2\theta_{23}){\rm sin}\,\theta_{13}\left[{\rm sin}(\Delta_{32}t)e^{-\gamma_{67} t}+{\rm sin}(\Delta_{21} t)e^{-\gamma_{12} t}-{\rm sin}(\Delta_{31} t)e^{-\gamma_{45} t}\right]\,, \\[3mm]
\Delta CP_{e\tau}=&P_{\nu_{e}\rightarrow \nu_{\tau}}-P_{\bar{\nu}_{e}\rightarrow \bar{\nu}_{\mu}}
=    -\Delta CP_{e\mu}\,, \\[3mm]
\Delta CP_{\mu\tau}=&P_{\nu_{\mu}\rightarrow \nu_{\tau}}-P_{\bar{\nu}_{\mu}\rightarrow \bar{\nu}_{\tau}}
=\Delta CP_{e\mu}\,.
\end{array}
\label{cp-violation}
\end{equation}
Note that the sum of $CP$ violations for fixed family is vanishing, as expected, i.e. we have
\begin{equation}
\Delta CP_{e\mu}+\Delta CP_{e \tau}= 0\,,\quad \Delta CP_{\mu e}+\Delta CP_{\mu \tau}= 0\,,\quad \Delta CP_{\tau e}+\Delta CP_{\tau \mu}= 0\,.\label{sum-zero}
\end{equation}
Moreover, they not depend on  $\gamma_3$ and $\gamma_8$.
Similar behaviors also manifest for $T$ violating channels:
\begin{equation}
\begin{array}{rl}
\Delta T_{e\mu}=&P_{\nu_{e}\rightarrow \nu_{\mu}}-P_{\nu_{\mu}\rightarrow \nu_{e}}=\Delta CP_{e\mu} \\[3mm]
\Delta T_{e\tau}=&P_{\nu_{e}\rightarrow \nu_{\tau}}-P_{\nu_{\tau}\rightarrow \nu_{e}}=\Delta CP_{e\tau} \\[3mm]
\Delta T_{\mu\tau}=&P_{\nu_{\mu}\rightarrow \nu_{\tau}}-P_{\nu_{\tau}\rightarrow \nu_{\mu}}=\Delta CP_{\mu\tau}\,.
\end{array}
\label{T-violation}
\end{equation}
Hence, in presence of a diagonal dissipation matrix, $CP$ and $T$ are violated, but $CPT$ is still preserved as in the standard case where no decoherence effects are present, i.e. $\Delta CP_{ab}=\Delta T_{ab}.$

Furthermore, it is clear that in such a case, the  Majorana phases $\phi_1$ and $\phi_2$ do not play any role, indeed all oscillation formula are independent of them. The violation of $CP$ and $T$ is related only to the presence of the Dirac phase, indeed if we set $\delta=0$ we recover $CP$ and $T$ invariance also in presence of a diagonal dissipator.
Different results are obtained for diagonal dissipators with  $\gamma_1 \neq \gamma_2$, or $ \gamma_4 \neq \gamma_5$, or $ \gamma_6 \neq \gamma_7$.
In these cases, one can show that the oscillation formulae and the $CP$ and $T$ violations depend on the Majorana phases.

\section{Non-diagonal dissipator}\label{non-diag dissip}

We now study the scenario with a non--diagonal dissipator.   We  consider the cases for which only two symmetric off--diagonal elements are non--zero.
In particular, we  mainly focus on following form for the dissipator:
\begin{equation}
D_{\mu0}=D_{0\nu}=0, \quad D_{11}= D_{22}=-\gamma_{12}, \quad D_{33}=-\gamma_{3},\quad D_{44}= D_{55}=-\gamma_{45},
 \quad D_{66}= D_{77}=-\gamma_{67}, \quad \quad D_{88}=-\gamma_8\,,\label{off-diag-diss-1}
\end{equation}
and then we will also comment on what happens if other off--diagonal elements are switched on. In the case described by Eq.\eqref{off-diag-diss-1} the system of differential equation in Eq.\eqref{syst-eq-diag} will differ for the components $\dot{\rho}_1$ and $\dot{\rho}_2$ which now satisfy the two differential equations
\begin{equation}
\begin{array}{rl}
\dot{\rho}_1(t)=&\displaystyle \Delta_{21}\rho_2(t)-\gamma_{12} \rho_1(t)-\alpha_1\rho_2(t)\,, \\[3mm]
\dot{\rho}_2(t)=&\displaystyle  -\Delta_{21}\rho_1(t)-\gamma_{12} \rho_2(t)-\alpha_1\rho_1(t)\,.
\end{array}
\label{syst-eq-non-diag}
\end{equation}
respectively, and whose solutions read
\begin{equation}
\begin{array}{rl}
\rho_{1}(t)=&\displaystyle e^{-\gamma_{12} t}\left[\rho_1(0){\rm cosh}(\Omega t)+\rho_{2}(0){\rm sinh}(\Omega t)\frac{\Xi_+}{\Omega}\right]\,, \\[3mm]
\rho_{2}(t)=&\displaystyle  e^{-\gamma_{12} t}\left[\rho_1(0){\rm sinh}(\Omega t)\frac{\Xi_{-}}{\Omega}+\rho_{2}(0){\rm cosh}(\Omega t)\right]\,,
\end{array}
\label{sol-diag-non-diag-alpha}
\end{equation}
while the other components are the same the ones in \eqref{sol-diag}. We have defined the quantities $\Omega\equiv \sqrt{\alpha_1^2-\Delta_{21}^2}$ and $\Xi_{\pm}\equiv \alpha_1\pm \Delta_{21}\,.$ The initial conditions $\rho_i(0)$ are the same as in Eqs. \eqref{sol-diag-electronic}, \eqref{sol-diag-muonic} and \eqref{sol-diag-tau} for electronic, muon  and tau  neutrinos, respectively.

Let us now distinguish two cases: (A) first, we consider a mixing matrix with zero Majorana phases to show the role played by  the Dirac phase in the violation of $CP$ and $CPT$ symmetries; (B) subsequently, we   compute   the   oscillation probabilities considering non--zero Majorana phases and analyze the effects on $CP$ and $CPT$ violations.

\subsection{Zero Majorana phases }

We set $\phi_1 = \phi_2 = 0,$ which means that we work with the mixing matrix $\mathcal{U}_D$ in Eq.\eqref{CKM}, i.e. with Dirac neutrinos.
We have the following results for the transitions preserving the flavor:
\begin{eqnarray}
\Delta CP_{ee}=&&\nonumber P_{\nu_e\rightarrow \nu_e}-P_{\bar{\nu}_{e}\rightarrow \bar{\nu}_e}=0\,,\\[3mm]\nonumber
\Delta CP_{\mu\mu}=&&P_{\nu_{\mu}\rightarrow \nu_{\mu}}-P_{\bar{\nu}_{\mu}\rightarrow \bar{\nu}_{\mu}}\\[3mm]\nonumber
=&&\displaystyle \frac{2\alpha_1 e^{-\gamma_{12} t}{\rm sin}\delta}{\Omega}{\rm sin}(2\theta_{23}){\rm sin}\,\theta_{13}{\rm sinh}(\Omega t)\left[{\rm cos}^2(\theta_{23}){\rm sin}(2\theta_{12}) \right.\\[3mm]\nonumber
&&\displaystyle\left.+{\rm sin}\,\theta_{13} \left({\rm cos}(2\theta_{12}){\rm sin}(2\theta_{23})-{\rm sin}(2\theta_{12}){\rm sin}^2\theta_{23}{\rm sin}\,\theta_{13}\right)\right] \,,\\[3mm]\nonumber
\Delta CP_{\tau\tau}=&&P_{\nu_{\tau}\rightarrow \nu_{\tau}}-P_{\bar{\nu}_{\tau}\rightarrow\bar{\nu}_{\tau}}\\[3mm]\nonumber
=&&\displaystyle \frac{2\alpha_1 e^{-\gamma_{12} t}{\rm sin}\delta}{\Omega}{\rm sin}(2\theta_{23}){\rm sin}\,\theta_{13}{\rm sinh}(\Omega t)\left[{\rm cos}^2(\theta_{23}){\rm sin}^2\theta_{13}{\rm sin}(2\theta_{12}) \right.\\[3mm]\nonumber
&&\displaystyle\left.-{\rm sin}(2\theta_{12}){\rm sin}^2\theta_{23}+{\rm cos(2\theta_{12}){\rm sin}\,\theta_{13}{\rm sin}(2\theta_{23})}\right] \,.\\[3mm]
\label{neutrino-antineutrino-non-diag}
\end{eqnarray}
In Eqs.(\ref{neutrino-antineutrino-non-diag}) it is shown that the violation of $CP$ appears in the transitions $\nu_{\mu}\rightarrow \nu_{\mu}$ and $\nu_{\tau}\rightarrow \nu_{\tau}.$ On the contrary, the transition $\nu_{e}\rightarrow \nu_{e}$ preserves such a  symmetry.
Notice that $\Delta CP_{\mu\mu}$ and $\Delta CP_{\tau\tau}$  does not appear either in absence of decoherence or in presence of a diagonal dissipator. As we will see in the next subsection, for Majorana neutrinos $\Delta CP_{ee} \neq 0$. Therefore,  the analysis  of such a violation  could be crucial in order to discriminate between Dirac and Majorana neutrinos in presence of an off--diagonal dissipation matrix.

Moreover, the $CP$ violating channels for different neutrinos are modified as follows:
\begin{equation}
\begin{array}{rl}
\Delta CP_{e\mu}=&P_{\nu_{e}\rightarrow \nu_{\mu}}-P_{\bar{\nu}_{e}\rightarrow \bar{\nu}_{\mu}} \\[3mm]
=&  \displaystyle  \frac{{\rm sin}\,\delta}{\Omega}{\rm cos}^2\theta_{13}{\rm sin}(2\theta_{12}){\rm sin}(2\theta_{23}){\rm sin}\,\theta_{13}\,\left[\Omega \left(e^{-\gamma_{45}t}{\rm sin}(\Delta_{31}t)-e^{-\gamma_{67}t}{\rm sin}(\Delta_{32} t)\right)\right.\\[3mm]
&\left.\displaystyle -e^{-\gamma_{12}t}(\alpha_{1}-\Delta_{21}){\rm sinh}(\Omega t)\right] \,, \\[3mm]
\Delta CP_{e\tau}=&P_{\nu_{e}\rightarrow \nu_{\tau}}-P_{\bar{\nu}_{e}\rightarrow \bar{\nu}_{\tau}}
= -  \Delta CP_{e\mu}\,, \\[3mm]
\Delta CP_{\mu\tau}=&P_{\nu_{\mu}\rightarrow \nu_{\tau}}-P_{\bar{\nu}_{\mu}\rightarrow \bar{\nu}_{\tau}} \\[3mm]
=&   \displaystyle \frac{{\rm sin}\,\delta\,}{4\Omega} {\rm sin}\,\theta_{13}{\rm sin}(2\theta_{23})\left[4\Omega {\rm cos}^2\theta_{13}{\rm sin}(2\theta_{12})\left(-e^{-\gamma_{67}t}{\rm sin}(\Delta_{32}t)+e^{-\gamma_{45}t}{\rm sin}(\Delta_{31} t)\right)\right.\, \\[3mm]
& \displaystyle -2e^{-\gamma_{12}t}\left({\rm sin}(2\theta_{12})\left(2\Delta_{21}{\rm cos}^2(\theta_{13})-\alpha_1\left({\rm cos}(2\theta_{13})-3\right){\rm cos}(2\theta_{23})\right)\right.\\[3mm]
& \left.\left.\displaystyle +4\alpha_1 {\rm cos}\,\delta {\rm cos}(2\theta_{12}){\rm sin}\,\theta_{13}{\rm sin}(2\theta_{23})\right) {\rm sinh}(\Omega t)\right] \,.
\end{array}
\label{cp-violation-alpha}
\end{equation}
%
%
The $T$ violations also differ from the diagonal case and are given by
\begin{equation}
\begin{array}{rl}
\Delta T_{e\mu}=&P_{\nu_{e}\rightarrow \nu_{\mu}}-P_{{\nu}_{\mu}\rightarrow \nu_{e}} \\[3mm]
=&   \displaystyle \frac{{\rm sin}\,\delta}{\Omega}\, {\rm cos}^2\theta_{13}{\rm sin}(2\theta_{12}){\rm sin}(2\theta_{23}){\rm sin}\,\theta_{13}\left[\Omega\left(e^{-\gamma_{45}t}{\rm sin}(\Delta_{31}t)-e^{-\gamma_{67}t}{\rm sin}(\Delta_{32} t)\right)-2\Delta_{21}e^{-\gamma_{12}t}{\rm sinh}(t\Omega)\right]\,, \\[3mm]
\Delta T_{e\tau}=&P_{\nu_{e}\rightarrow \nu_{\tau}}-P_{\nu_{\tau}\rightarrow \nu_{e}}
=    -\Delta T_{e\mu}\,, \\[3mm]
\Delta T_{\mu\tau}=&P_{\nu_{\mu}\rightarrow \nu_{\tau}}-P_{\nu_{\tau}\rightarrow \nu_{\mu}}
=\Delta T_{e\mu}\,.
\end{array}
\label{T-viol-dirac}
\end{equation}
Therefore, unlike the case of a diagonal dissipator, when $\alpha_1\neq 0$, not only $CP$ and $T$ are violated, but also $CPT$ symmetry is not preserved:
\begin{equation}
\Delta CP_{e\mu}  \neq \Delta T_{e\mu}\,,\qquad  \Delta CP_{e\tau} \neq \Delta T_{e\tau}\,,\qquad  \Delta CP_{\tau\mu}\neq \Delta T_{\tau\mu}\,.
\end{equation}
Such violations  are related to the presence of the Dirac phase, indeed by setting $\delta=0$, all the three symmetries are preserved even if $\alpha_1\neq 0.$ Let us point out that such an effect is not present in the two flavors case analyzed in \cite{Capolupo:2018hrp} since in that case no Dirac phase is present and one can not find any relation between the phase $\delta$ and $CPT$ violation. The $CPT$ violation induced by Dirac phase is a new feature in presence of decoherence and dissipation. If we set $\alpha_1=0$ we recover the case of diagonal dissipator where $CPT$ symmetry is preserved. Furthermore, in presence of an off--diagonal dissipator, the oscillation formula depends on the choice of the mixing matrix, indeed one can straightforwardly check that different parametrizations of the mixing matrix for Dirac neutrinos give different physical results. In this paper we focus on the Pontecorvo-Maki-Nakagawa (PMNS) parametrization in Eq.\eqref{CKM} for Dirac neutrinos. Our results show that, if the decoherence characterizes neutrino oscillations,   next long baseline experiments could reveal which matrix elements contain the $\delta$--phase.

Let us emphasize that so far we have only considered one possible case of non--diagonal dissipator, in which only $\alpha_1$ is non--zero. Of course, also other kinds of dissipation matrices can be studied in which other off--diagonal elements are non--zero.
By making computations similar to those presented above, one can show that  all the possible choices of the dissipator \eqref{dissip-matrix} lead to $CP$ and $T$  violations, as it also happens in the diagonal case. On the other hand, $CPT$ is violated in most of the cases; however, there are some off--diagonal choices which still preserve it. Indeed, $CPT$ symmetry is   respected when the only non--zero off--diagonal element is one of the following: $\beta_1,\,\alpha_3,\,\delta_3,\,\xi_3,\,\eta_1,\,\zeta_2,\,\chi_4,\,\delta_5,\,\beta_6,\,\alpha_7,\,\gamma_8\,.$

\subsection{Non--zero Majorana phases }

In this Subsection we repeat the previous analysis for the mixing matrix in Eq.\eqref{majorana-matrix} where the Majorana phases $\phi_1$ and $\phi_2$ are non--zero. We show that in presence of an off--diagonal dissipator, the oscillation formulae, the $CP$ and $T$ violations can depend on the Majorana phases, thus providing a new framework in which the real nature of neutrino can be challenged.

By working with the dissipator in Eq.\eqref{off-diag-diss-1} and using the parametrization in Eq.\eqref{majorana-matrix},  we obtain the following $CP$ violations for the transitions preserving the flavor:
\begin{eqnarray}
\Delta CP^M_{ee}=\nonumber&&P^M_{\nu_e\rightarrow \nu_e}-P^M_{\bar{\nu}_{e}\rightarrow \bar{\nu}_e}\label{e-e-majo}\\[3mm]
=&&\displaystyle -\frac{2\alpha_1 e^{-\gamma_{12} t}{\rm sin}\,\delta}{\Omega}{\rm sin }(2\phi_1){\rm cos}^4\theta_{13}{\rm sin}^2(2\theta_{12}){\rm sinh}(\Omega t)  \,,\\[3mm]\nonumber
\Delta CP^M_{\mu\mu}=\nonumber&&P^M_{\nu_{\mu}\rightarrow \nu_{\mu}}-P^M_{\bar{\nu}_{\mu}\rightarrow \bar{\nu}_{\mu}}\\[3mm]\nonumber
=&&\displaystyle \frac{2\alpha_1 e^{-\gamma_{12} t}{\rm sin}\,\delta}{\Omega}\left[{\rm cos}\,\phi_1\,{\rm cos}^2\theta_{23}{\rm sin}(2\theta_{12})-{\rm sin}\,\theta_{13}\left({\rm cos}\,\phi_1\,{\rm sin}^2(2\theta_{23}){\rm sin}\,\theta_{13}{\rm sin}(2\theta_{12})\right.\right.\\[3mm]\nonumber
&& \displaystyle \left.\left.+ {\rm sin}(2\theta_{23})\left({\rm cos}(\delta+\phi_1){\rm sin}^2\theta_{12}-{\rm cos}(\delta-\phi_1){\rm cos}^2\theta_{12} \right) \right)\right]\\[3mm]\nonumber
&&\displaystyle \times \left[ {\rm cos}^2\theta_{12}{\rm sin}\,\theta_{13}\,{\rm sin}(2\theta_{23}){\rm sin}(\delta-\phi_1)-{\rm cos}^2\theta_{23}{\rm sin}(2\theta_{12}){\rm sin}\,\phi_1     \right.\\[3mm]\label{mu-mu-majo}
&&\displaystyle\left.+{\rm sin}\,\theta_{13}\left({\rm sin}\,\phi_1\,{\rm sin}^2\theta_{23}{\rm sin}(2\theta_{12}){\rm sin}\,\theta_{13}+{\rm sin}^2\theta_{12}{\rm sin}(2\theta_{23}){\rm sin}(\delta+\phi_1)\right)\right] {\rm sinh}(\Omega t) \,,\\[3mm]\nonumber
\Delta CP^M_{\tau\tau}=\nonumber&&P^M_{\nu_{\tau}\rightarrow \nu_{\tau}}-P^M_{\bar{\nu}_{\tau}\rightarrow \bar{\nu}_{\tau}}\\[3mm]\nonumber
=&&\displaystyle \frac{2\alpha_1 e^{-\gamma_{12} t}{\rm sin}\,\delta}{\Omega}\left[{\rm cos}\,\phi_1\,{\rm cos}^2\theta_{23}{\rm sin}(2\theta_{12}){\rm sin}^2\theta_{13}-{\rm cos}\,\phi_1\,{\rm sin}(2\theta_{12}){\rm sin}^2\theta_{23}\right.\\[3mm]\nonumber
&& \displaystyle \left.+ {\rm sin}(2\theta_{23}){\rm sin}\,\theta_{13}\left({\rm cos}(\delta-\phi_1){\rm cos}^2\theta_{12}-{\rm cos}(\delta+\phi_1){\rm sin}^2\theta_{12} \right)\right]\\[3mm]\nonumber
&&\displaystyle \times \left[ {\rm cos}^2\theta_{12}{\rm sin}\,\theta_{13}\,{\rm sin}(2\theta_{23}){\rm sin}(\delta-\phi_1)-{\rm cos}^2\theta_{23}{\rm sin}^2\theta_{13}{\rm sin}(2\theta_{12}){\rm sin}\,\phi_1     \right.\\[3mm]\label{tau-tau-majo}
&&\displaystyle\left.+{\rm sin}\,\phi_1\,{\rm sin}^2\theta_{23}{\rm sin}(2\theta_{12})+{\rm sin}^2\theta_{12}{\rm sin}(2\theta_{23}){\rm sin}\,\theta_{13}{\rm sin}(\delta+\phi_1)\right]{\rm sinh}(\Omega t) \,.
\end{eqnarray}
Here,  with the letter $M$ we mean the transition probabilities for Majorana neutrinos. By comparing Eqs.\eqref{e-e-majo}, \eqref{mu-mu-majo} and \eqref{tau-tau-majo} with the analogue in Eq.\eqref{neutrino-antineutrino-non-diag}, we can immediately note that the presence of non--zero Majorana phases introduces new terms in the formulae, and in particular, generate a $CP$ violation also in the transition $\nu_e\rightarrow \nu_e.$ This violation is absent  for Dirac neutrinos, and depends on $\phi_1 $ for the dissipator considered.

The transition \eqref{e-e-majo} has a very peculiar meaning: unlike the case of zero Majorana phases, here  $\Delta CP_{ee}$ turns out to be non--vanishing, and becomes zero only when $\phi_1=0\,.$ Such a feature is crucial in order to discriminate between Dirac and Majorana neutrinos and provides a completely new way to test the real nature of neutrinos in future experiments. Indeed, by considering the mixing matrix in Eq.\eqref{majorana-matrix} and the dissipator in Eq.\eqref{off-diag-diss-1}, we have $\Delta CP_{ee}=0$ for Dirac neutrinos and $\Delta CP^M_{ee}\neq 0$ for Majorana neutrinos. Let us also clarify that such a difference in the $CP$ violation for $\nu_{e}\rightarrow \nu_{e}$ transition, with respect to the other two, depends on the form of the dissipation matrix and on the representation of the mixing matrix for Majorana neutrinos.

The possibility to violate the $CP$ symmetry in the transitions flavor preserving, here revealed, is a new result which can indicate the presence of decoherence and allow us to fix the form of the mixing matrix, besides the neutrino nature.

The $CP$ violations for transitions between  different neutrinos are:
\begin{eqnarray}\nonumber
\Delta CP_{e\mu}^M=&&P^M_{\nu_{e}\rightarrow \nu_{\mu}}-P^M_{\bar{\nu}_{e}\rightarrow \bar{\nu}_{\mu}} \\[3mm]\nonumber
=&&   -\displaystyle \frac{{\rm cos}\theta_{13}}{2\Omega}\left[\Omega{\rm sin}\,\delta\,\left(-e^{-\gamma_{67}t}{\rm sin}(\Delta_{32}t)+e^{-\gamma_{45}t}{\rm sin}(\Delta_{31} t)\right){\rm sin}(2\theta_{12}){\rm sin}(2\theta_{13}){\rm sin}(2\theta_{23})
\right. \\[3mm] \nonumber
&& \displaystyle +\frac{e^{-\gamma_{12}t}}{2}\left(\alpha_1{\rm cos}\theta_{13}\left(2{\rm cos}^2\theta_{13}-({\rm cos}(2\theta_{13})-3){\rm cos}(2\theta_{23})\right){\rm sin}(2\phi_1){\rm sin}^2(2\theta_{12}) \right.\\[3mm] \nonumber
&& \displaystyle\left.\left. +\left(-2(\Delta_{21}+\alpha_1{\rm cos}(2\phi_1)){\rm cos}\,\delta {\rm sin}(4\theta_{12})\right){\rm sin}(2\theta_{13}){\rm sin}(2\theta_{23})\right){\rm sinh}( \Omega t)\right]\,, \label{cp-violation-alpha-majo-mu-e}
\\[3mm]
\Delta CP^M_{e\tau} =&&P^M_{\nu_{e}\rightarrow \nu_{\tau}}-P^M_{\bar{\nu}_{e}\rightarrow \bar{\nu}_{\tau}}= - \Delta CP_{e\mu}^M\,,
\label{cp-violation-alpha-majo-e-tau}
\end{eqnarray}
We do not report explicitly the expression of $\Delta CP^{M}_{\mu\tau}$ because of its length. Its behavior is depicted in the left panel of Fig.\ref{fig1}.

The $T$ violating channels are not affected by the Majorana phases for our choice of the dissipator, indeed they are the same as in Eq.\eqref{T-viol-dirac}:
\begin{equation}
\Delta T^M_{e\mu}=\Delta T_{e\mu}\,,\quad \Delta T^M_{e\tau}=\Delta T_{e\tau}\,,\quad \Delta T^M_{\mu\tau}=\Delta T_{\mu\tau}\,.
\end{equation}
This fact induces  an extra violation of the $CPT$ symmetry since we have
\begin{equation}
\Delta CP^M_{e\mu}\neq \Delta T_{e\mu}\,,\qquad \Delta CP^M_{e\tau}\neq\Delta T_{e\tau}\,,\qquad \Delta CP^M_{\tau\mu}\neq \Delta T_{\tau\mu}\,.
\end{equation}
In presence of an off--diagonal dissipator, Dirac and Majorana phases induce two independent $CPT$ violations.
The results here presented are obtained by considering the non--diagonal dissipator in which only $\alpha_1$ is non--zero; see Eq.\eqref{off-diag-diss-1}. Other kinds of dissipation matrices can be studied with other off--diagonal elements switched on.
Like for the mixing matrix in Eq.\eqref{CKM}, also for the matrix \eqref{majorana-matrix}, $CP$ and $T$ are always violated, while $CPT$ can be still preserved for some non--zero off--diagonal elements. Indeed, $CPT$ is respected if the only non--zero off--diagonal element is one of among these: $\beta_1,\,\alpha_3,\,\delta_3,\,\xi_3,\,\eta_1,\,\zeta_2,\,\chi_4,\,\delta_5,\,\beta_6,\,\alpha_7,\,\gamma_8\,.$

Notice also that other choices of the Majorana matrix would give different results. For instance the mixing matrix $\mathcal{U}_M$ in Eq.\eqref{non-interest-majo}  give different expressions for the oscillation formula as compared to Eq.\eqref{majorana-matrix}. This implies that the physical results depend on the chosen parametrization of the Majorana mixing matrix.

Summarizing,  in presence of an off--diagonal dissipator, the neutrino oscillation formula depend on the   parametrization of the mixing matrix. A physical implication is that Dirac and Majorana neutrinos are two totally distinct entities and their nature, together with $CPT$ violation, can be tested with future experiments.

\section{Comparison between Dirac and Majorana neutrinos}\label{numeric-analysis}

In this Section we  relate our theoretical analysis to the parameters  of  neutrino experiments. We compare the behavior of Dirac and Majorana neutrinos considering some specific transition probabilities.
In order to connect our results with existing long baseline experiments such as IceCube  and DUNE, one should consider neutrino propagation in the matter and to adopt the formalism presented in Ref.\cite{Carpio:2017nui}, which generalize the Mikheyev-Smirnov-Wolfenstein (MSW) effect \cite{MSW1,MSW2,MSW3}  to the case of decoherence.
However, since the Earth is not charge-symmetric (it contains electrons, protons and neutrons, but it does not contain their antiparticles),
then the oscillations in matter involving electron neutrino
already induce  the $CP$ and $CPT$ violations also in absence of decoherence. Therefore, one has to be careful to identify the right contribution responsible for violations purely induced by decoherence. Since we are mainly interested in highlighting the effects of the decoherence,  we consider the neutrino oscillations in vacuum. In the  following we approximate $x\approx t$ in Natural units.

\begin{figure}[t!]
	\centering
	\subfloat[Subfigure 1 list of figures text][]{
		\includegraphics[scale=0.39]{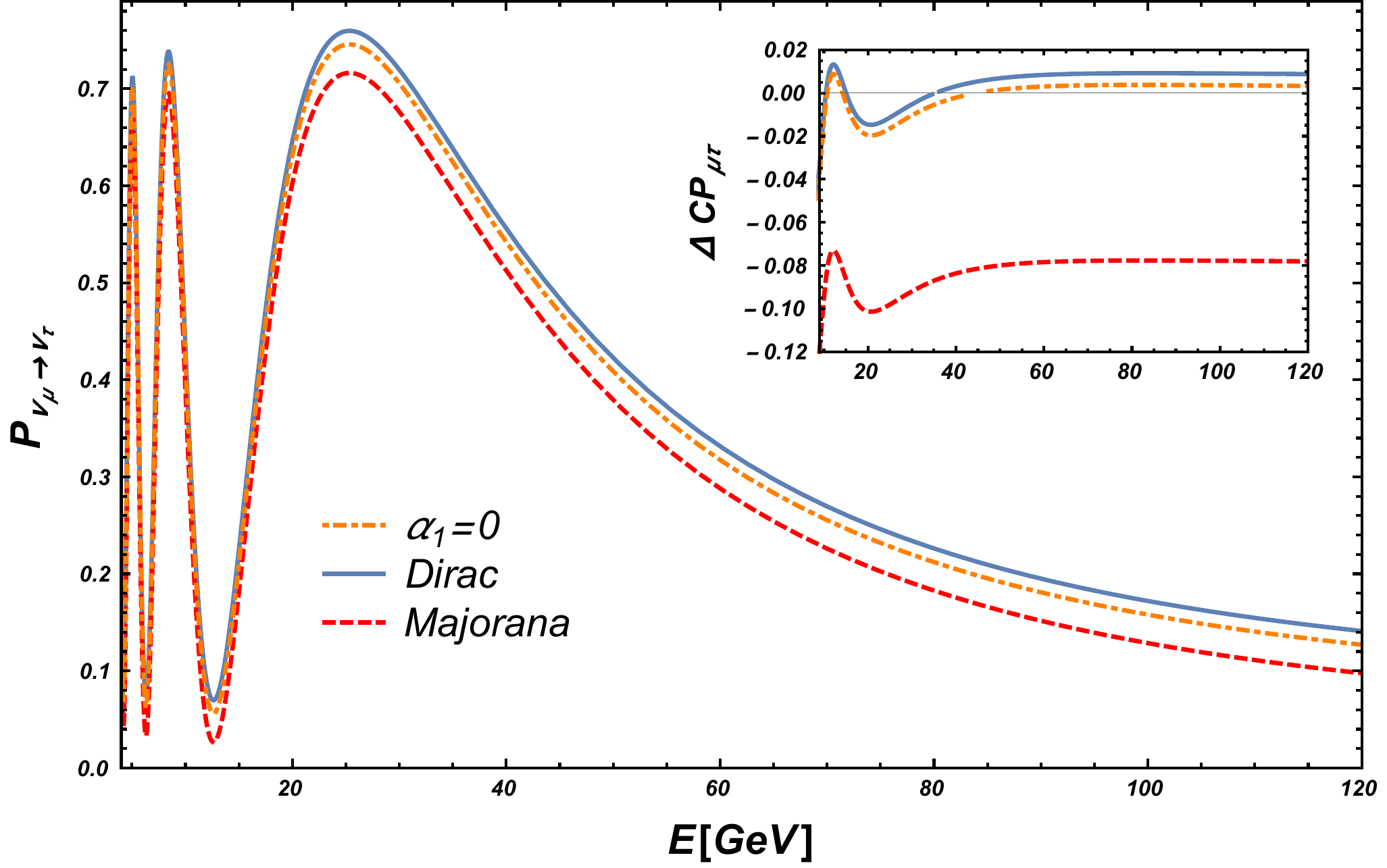}
		\label{fig.2.a}}
	\qquad
	\subfloat[Subfigure 2 list of figures text][]{
		\includegraphics[scale=0.385]{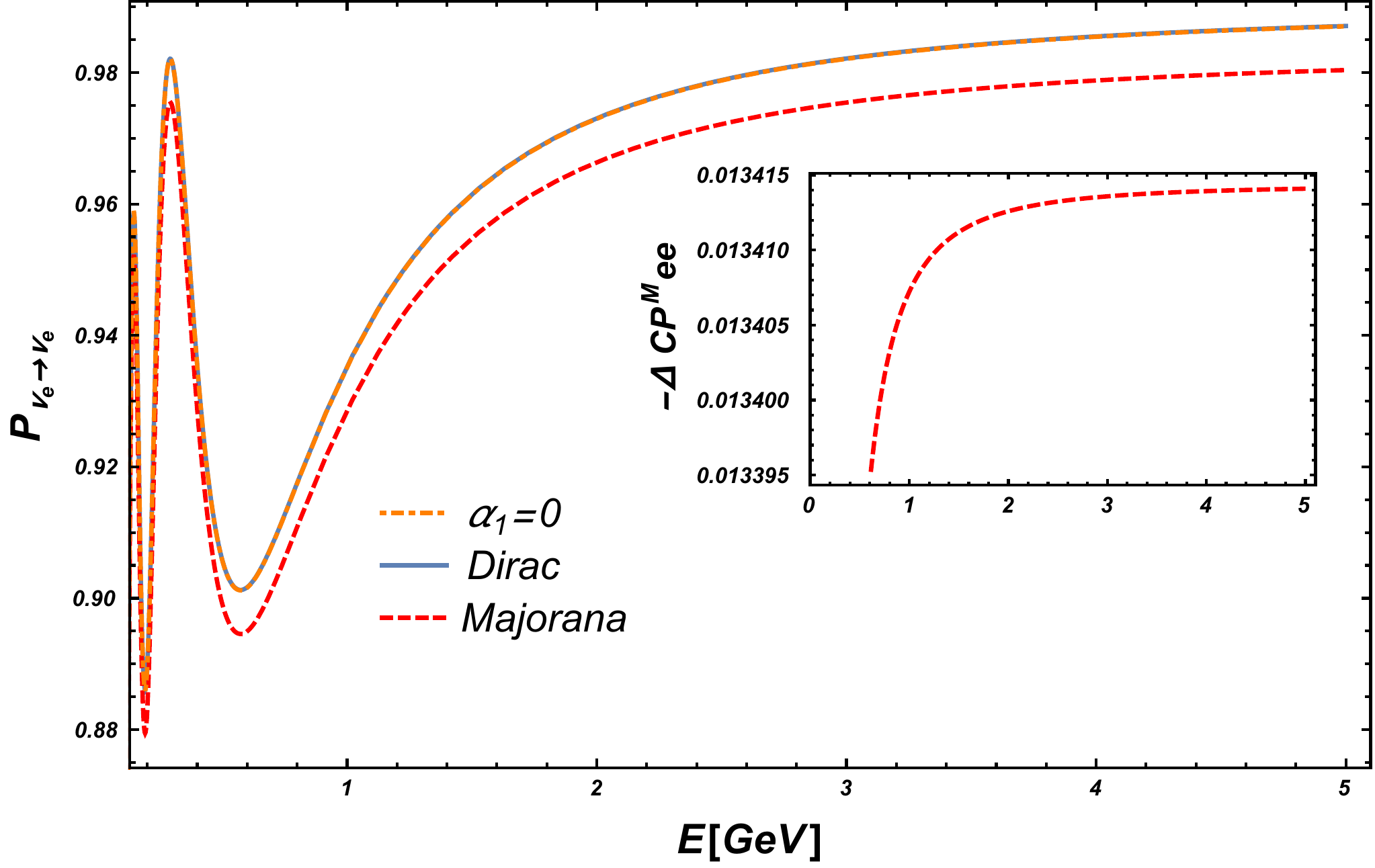}
		\label{fig.2.b}}
	\protect\caption{(a) Plots of the transition probability $P_{\nu_\mu\rightarrow\nu_\tau}$ as a function of the energy $E$ for a diagonal dissipator $\alpha_{1} = 0$, for which Dirac and Majorana neutrinos have identical behavior (orange dot--dashed line) and for the off--diagonal dissipator for   Dirac (blue solid line) and Majorana (red dashed line) neutrinos. Inset: corresponding plots of $\Delta CP_{\mu\tau}.$ We consider the energy range $(0-120)$GeV corresponding to the accessible energies in the IceCube DeepCore experiment and set $t=6.58\times 10^{22}{\rm GeV}^{-1}.$
\\
	(b) Plots of $P_{\nu_e\rightarrow \nu_e}$ as a function of the energy   for an off--diagonal dissipator in the case of Dirac (blue solid line) and Majorana (red dashed line) neutrinos. Notice that for this transition, the behavior of Dirac neutrinos in the off--diagonal dissipator case is identical to that of the neutrinos in the case of a diagonal dissipator. Inset: plot  of the $CP$ violation in the channel $\nu_e\leftrightarrow \nu_e$ for Majorana neutrinos. We consider the energy values $(0.3-5)$GeV characteristic of DUNE experiment and we set $t=1.49\times 10^{21}{\rm GeV}^{-1}.$ In both the plots we assume  $\phi_1 = \pi/4$, $\phi_2=\pi/3,$ $\delta=-\pi/2,$  we consider the following values of  the elements of the dissipator: $\gamma_{12}=1.2 \times 10^{-23}{\rm GeV}$, $\gamma_{45}=4 \times 10^{-24}{\rm GeV}$, $\gamma_{67}=4.7 \times 10^{-24}{\rm GeV}$, $\gamma_3=\gamma_8=7.9 \times 10^{-24}{\rm GeV},$ $\alpha_1 = 1.3 \times 10^{-24}{\rm GeV}, $ and use the following experimental values for the mixing angles:  $\sin^{2} \theta_{23} = 0.51 $, $\Delta m^{2}_{23} = 2.55 \times 10^{- 3} {\rm eV}^{2},$  $\Delta m^{2}_{12} = 7.56 \times 10^{- 5} {\rm eV}^{2}$.}\label{fig1}
\end{figure}

In Fig. \ref{fig1}, panel (a), we plot the $\nu_\mu \rightarrow \nu_\tau$ oscillations in vacuum  and  $\Delta CP_{\mu\tau}$ as functions of the neutrino energy, by using the range of energy of the IceCube DeepCore experiment $E\in(6-120)$GeV \cite{Aartsen:2017nmd,Coloma} and  a distance equal to Earth diameter $ x = 1.3\times 10^{4} $km, corresponding to   $t=6.58\times 10^{22}{\rm GeV}^{-1}$. We draw the oscillation formula $P_{\nu_{\mu} \rightarrow \nu_\tau}$ and the quantity $\Delta CP_{\mu\tau}$ obtained by using the diagonal and the off--diagonal dissipators with zero and non--zero Majorana phases, respectively.

In panel (b), we plot the oscillation formula  $P_{\nu_e \rightarrow \nu_e}$ and $\Delta CP_{e e }$ in the energy range $(0.3-5)$GeV which is typical of DUNE experiment \cite{Balieiro}. We consider the time scale $t=1.49\times 10^{21}{\rm GeV}^{-1}.$
For both the plots, we assume  $\phi_1 = \pi/4$, $\phi_2=\pi/3,$ $\delta=-\pi/2;$ and we use the following values for the elements of the dissipator: $\gamma_{12}=1.2 \times 10^{-23}{\rm GeV}$, $\gamma_{45}=4.0 \times 10^{-24}{\rm GeV}$, $\gamma_{67}=4.7 \times 10^{-24}{\rm GeV}$, $\gamma_3=\gamma_8=7.9 \times 10^{-24}{\rm GeV},$ $\alpha_1 = 1.3 \times 10^{-24}{\rm GeV},$ which are compatible with the experimental upper bounds on $\gamma_i$ \cite{Coloma,Balieiro}. Moreover, we consider the following experimental values of the parameters:  $\sin^{2} \theta_{23} = 0.51 $, $\Delta m^{2}_{23} = 2.55 \times 10^{- 3} {\rm eV}^{2},$  $\Delta m^{2}_{12} = 7.56 \times 10^{- 5} {\rm eV}^{2}$ \cite{Coloma}.

The plots show different behaviors between Dirac and Majorana neutrinos, induced by decoherence which could be detected in next long baseline experiments.
In our treatment we have neglected the effect of matter,
we leave  for future works  a detailed investigation on matter effects on $CP$ and $CPT$ violations in the presence of decoherence.

\section{Summary and conclusions}\label{conclus}

In this work we have analyzed the physical implications of decoherence and dissipation in the context of three flavors neutrino mixing. We have computed the transition probabilities for Dirac and Majorana neutrinos  in   the cases of a diagonal and an off--diagonal dissipation matrix. By analyzing Dirac neutrinos, we have shown that in presence of a diagonal dissipator, the oscillation formula do not depend on the   parametrization of the mixing matrix and $CPT$ symmetry is still preserved. Subsequently,   we have switched on an off--diagonal elements in the dissipation matrix, and shown that for Dirac neutrinos the oscillation formula can depend on the parametrization   of the mixing matrix. Moreover, we have revealed the possibility of a $CP$ violation in the neutrino transitions preserving the flavor and the existence of a $CPT$ violation due to the Dirac phase $\delta\,.$
By performing analogue computations for Majorana neutrinos, we have  shown that in presence of an off--diagonal dissipation matrix, the oscillation formulae can depend on the Majorana phases $\phi_i$. These formulae depend on the  choices of the parametrization of the Majorana mixing matrix. Indeed, different parametrizations lead to different formulae.  We have also revealed a $CPT$ violation term purely induced by the Majorana phases, which generalize the result in \cite{Capolupo:2018hrp} obtained for two flavors neutrinos.

For a specific form of the dissipator whose non--zero off--diagonal element is $\alpha_1,$ we have shown that $\Delta CP_{ee}=P_{\nu_e\rightarrow \nu_e}-P_{\bar{\nu}_{e}\rightarrow \bar{\nu}_e} $  is zero for Dirac neutrinos, while it is non--vanishing for Majorana ones. $\Delta CP_{ee}$ could be analyzed in next experiments to discriminate between Dirac and Majorana neutrinos.

The $CPT$ violation induced by Dirac and Majorana phases, together with the distinction in the oscillation formula for Dirac and Majorana neutrinos, can be really tested in long baseline experiments if the phenomenon of decoherence is not negligible. Very interestingly, such a phenomenon might be even more accessible than the neutrinoless double beta decay, and represent a totally new scenario where to test the real nature of neutrinos. By using the parameters of IceCube DeepCore and DUNE experiments, and the constraints on dissipation matrix \cite{Coloma,Balieiro}, we have analyzed the transitions $\nu_{\mu}\rightarrow \nu_{\tau}$ and $\nu_{e}\rightarrow \nu_{e},$ and made a comparison between Dirac and Majorana.
A detection of $CPT$ violation induced by decoherence effects could be   attributed to fluctuations of the space-time \cite{Rovelli,Rovelli1}, thus such a detection might represent a  signature of quantum gravity. Moreover, the studies on neutrino mixing in curved space \cite{Cardall:1996cd,Buoninfante:2019der} could be also generalized by including in them the decoherence and dissipation effects here presented.
Therefore, our study might open new windows of opportunity to address   several   open questions in fundamental physics.
It is worthwhile note that, non-perturbative field theoretical
effects of particle mixing \cite{Blasone:1998hf}, \cite{Capolupo:2006et}  can be neglected in the our treatment.

\acknowledgments
L.B. acknowledges support from JSPS No.~P19324 and KAKENHI Grant-in-Aid for Scientific Research No.~JP19F19324. A.C. and G.L. acknowledges partial financial support from MIUR, INFN and COST Action CA1511 Cosmology and Astrophysics Network for Theoretical Advances and Training Actions (CANTATA). S.M.G. acknowledges support from the European Regional Development Fund the Competitiveness and Cohesion Operational Programme (KK.01.1.1.06--RBI TWIN SIN), the Croatian Science Fund Project No. IP--2016--6--3347 and IP--2019--4--3321  and the QuantiXLie Center of Excellence  (Grant KK.01.1.1.01.0004).


\end{document}